\documentclass[slac_one]{revtex4}
\usepackage{graphicx}
\usepackage{fancyhdr}
\begin{document}

\title{The Belle II Experiment at the SuperKEKB} 

%

\author{C. Z. Yuan (for the Belle II Collaboration)}
\affiliation{Institute of High Energy Physics, Chinese Academy of
Sciences, Beijing 100049, China}

\begin{abstract}

Belle II experiment at the SuperKEKB collider is a major upgrade
of the Belle experiment at the KEKB asymmetric $e^+e^-$ collider
at the KEK. The experiment will focus on the search for new
physics beyond the standard model via high precision measurement
of heavy flavor decays and search for rare signals. In this talk,
we present the status of the SuperKEKB collider and the Belle II
detector.

\end{abstract}

\maketitle

\thispagestyle{fancy}


\section{Introduction}

The two B factories, Belle at the KEKB collider at
KEK~\cite{Belle:2000cg} and BaBar at the PEP II collider at
SLAC~\cite{Aubert:2001tu}, have been operating very successfully
for more than ten years. Altogether, about 1.5~ab$^{-1}$ data in
the vicinity of the $\Upsilon(4S)$ resonance were accumulated, and
these data were used in extensive studies of the B-physics, charm
physics, tau physics, as well as hadron spectroscopy, and various
searches for possible hints of new physics, such as lepton number
violation, low mass Higgs-like states, and dark matters.

While most results from B factories are in good consistency with
the expectations from the Standard Model (SM), there are a few
measurements that show discrepancies at around three standard
deviation level from the SM predictions. One of the interesting
examples is the unexpected large branching fraction of pure
leptonic decay $B^+\to \tau^+\nu$, if confirmed, very probably,
there are contributions from non-SM intermediate
particles~\cite{ss,Aushev:2010bq}.

Much larger data sets are needed to investigate whether these are
hints for New Physics (NP) or merely statistical fluctuations. The
main purpose of a super flavor factory, SuperKEKB, is to
accumulate a much larger data set than at the B factories, to pin
down the experimental uncertainties, and to confirm or deny the
discrepancies from the SM predictions~\cite{Abe:2010sj}.

The KEKB collider has achieved world record luminosity of
$2.1\times 10^{34}~{\rm cm}^{-2}s^{-1}$ and the Belle experiment
accumulated slightly more than 1~ab$^{-1}$ in about 10 years. To
reach the physics goals of the next generation flavor-factory
experiment, both the accelerator and the detector need to be
upgraded significantly. The designed luminosity of the SuperKEKB
is $8\times 10^{35}~{\rm cm}^{-2}s^{-1}$ which is about 40 times
higher than the KEKB, and the expected integrated luminosity at
Belle II experiment is 50~ab$^{-1}$ in about 5 years
running~\cite{Abe:2010sj}.

\section{SuperKEKB Collider}

The luminosity at an $e^+e^-$ collider, $\mathcal{L}$, is given by
\begin{equation}
\mathcal{L} = \frac{\gamma_{e^\pm}}{2er_e} \left(1 +
\frac{\sigma_y^*}{\sigma_x^*} \right) \frac{I_{e^\pm}
\xi_y^{e^\pm}}{\beta_y^{*e^\pm}}
\left(\frac{R_L}{R_{\xi_y}}\right) \label{eqn:lumi}
\end{equation}
where $\gamma$ is the Lorentz factor, $\sigma_y^*/\sigma_x^*$ the
beam size aspect ratio, $I$ the beam current, $\beta^*_y$ the
vertical beta function at the interaction point, $\xi_y$ the
beam-beam parameter, and $R_L/R_{\xi_y}$ a geometrical factor. The
$e^\pm$ refers to the product of the corresponding quantities for
the low energy positron (LER) and high energy electron (HER)
beams.

In order to achieve a factor of 40 increase in the luminosity than
at KEKB, all the above parameters have to be improved. The main
increase in luminosity comes from a significantly smaller beam
size at the interaction point (the so-called nano-beam scheme).
The beta functions at the interaction point are reduced in $y$
direction from 5.9/5.9~mm to 0.27/0.31~mm for HER/LER, and in $x$
direction from 1200~mm to 32/25~mm.

The beam-beam parameter is proportional to
$\sqrt{\beta^*/\epsilon}$, the emittance $\epsilon$ is reduced to
keep the beam-beam parameter at the same level as at KEKB. A
reduction of the emittance from 18/24~nm to 3.2/(4.3-4.6)~nm is
obtained by installing a new electron source and a new damping
ring, in addition to a redesign of the HER arcs. The last
contribution to the luminosity gain comes from higher beam
currents. They are increased from 1.64/1.19~A to 3.60/2.60~A.

To improve the lifetime of the LER beam, the beam energies are
changed from 3.6/8~GeV to 4/7~GeV, and the crossing angle of the
two beams is increased from 22~mrad to 83~mrad.

\section{Belle II Detector}

\begin{figure}
\centering
\includegraphics[width=0.9\textwidth]{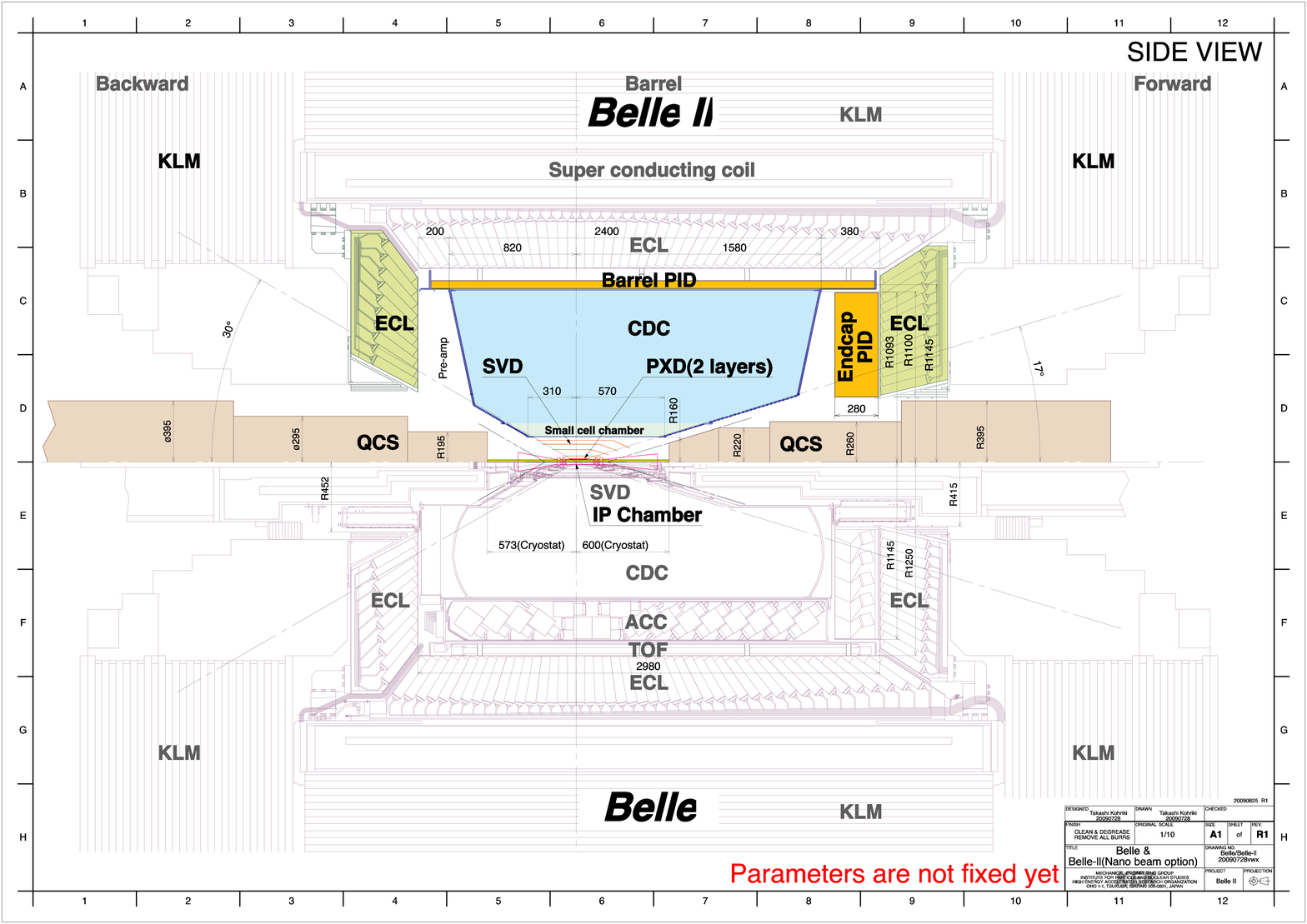}
\caption{The Belle II detector (top half) compared with the Belle
detector (bottom half).} \label{detector}
\end{figure}

The increased background level (10-20 times more than at Belle)
requires the Belle II detector to be able to handle the higher
occupancy and radiation damage than at Belle. The about a factor
of 10 increased event rate requires improved trigger, data
acquisition, and computing systems. To reach higher precision
measurement of some physics quantities requires improved low
momentum $\mu$ identification and good neutrino reconstruction via
missing energy. The components of the Belle detector are either
significantly upgraded or replaced with new ones.
Figure~\ref{detector} shows a comparison of the Belle and Belle II
detectors.

The most challenging experimental requirement is the detection of
the decay point of the short-living B-mesons, relying on a
high-performance vertex detector. The two inner layers of the
Belle II vertex detector will use two layers of silicon pixel
sensors (PXD) based on the DEPFET technology surrounded by four
layers of double sided silicon strip detectors (SVD). With the
excellent spatial resolution of the PXD, an impact parameter
resolution in beam direction of $\sim 10~\mu$m can be achieved.
The larger outer radius of the SVD compared to Belle gives an
increase in efficiency of about 30\% for the reconstruction of
$K_S \rightarrow \pi^+\pi^-$ decays inside the SVD. A precise
measurement of the momentum and the energy loss (dE/dx) of charged
tracks is provided by the central drift chamber (CDC) with a
higher granularity, a new gas - He/${\rm C}_2{\rm H}_6$ and an
improved readout system. It will start immediately after the
vertex detector and cover range 16~cm $< r <$ 112~cm of distances
from the beam axis.

Identification of hadrons is another key element for Belle II: it
has to separate kaons from pions with high efficiency and very low
fake rate. This will be provided by replacing the time-of-flight
detector at Belle with the Time-of-Propagation (TOP) counter in
the barrel region and the proximity focusing Cherenkov ring
imaging counter with aerogel radiators in the endcaps. The TOP
counter will measure the time that the internally reflected light
travels down the quartz bar, together with two spatial coordinates
of the photon impact point at the bar end surface. A kaon
identification efficiency of $>$97\% at a pion misidentification
rate of $<$1\% is expected for almost the full momentum and
angular ranges.

The electromagnetic calorimeter (ECL) will face a severe
background increase. While the crystals of the Belle
electromagnetic calorimeter will be reused for the barrel part,
the readout electronics will be replaced by new ones with pipeline
and waveform sampling. In the endcaps, pure CsI will be used
because it is more radiation tolerant and faster, thus better
coping with pile up problems.

Muons and $K_L$ mesons are identified by resistive plate chambers
in the outer part of the Belle detector (KLM). For Belle II the
endcap regions as well as the first three layers in the barrel
region will be upgraded with scintillator strips to cope with the
high background rates.

Computing system of the Belle II will use the newly developed
BASF2 framework. It will have to handle an amount of data
eventually corresponding to 50 times the Belle level in about 5
years' running. This means an amount of raw data of the order of
$10^{10}$ events per year. Therefore, a distributed computing
model based on the grid will be adopted.

\section{Summary}

The upgrade of the B-factory at KEK is in progress. The
accelerator, SuperKEKB, with a designed luminosity of $8 \times
10^{35}~{\rm cm}^{-2}s^{-1}$, will starts physics running by the
end of 2015, and will supply an integrated luminosity of
10~ab$^{-1}$ in 2018, and 50~ab$^{-1}$ in 2020. The Belle detector
components are either upgraded or replaced by new ones to cope
with the more challenging beam conditions and to improve the
detector performance. For more details on the physics and the
experiment, one may refer to Refs.~\cite{Aushev:2010bq}
and~\cite{Abe:2010sj}.

\section{Acknowledgements}

We congratulate the organizers for a great workshop. The work is
supported in part by National Natural Science Foundation of China
(NSFC) under Contracts Nos. 10825524 and 10935008.

\end{document}